\documentclass[11pt]{article}

\usepackage{amsmath,amssymb,mathtools,amsfonts,epsfig,graphicx,euscript}
\usepackage{amsmath,amssymb}
\newcommand{\bse}{\begin{subequations}}
\newcommand{\ese}{\end{subequations}}
\newcommand{\be}{\begin{equation}}
\newcommand{\ee}{\end{equation}}
\newcommand{\bea}{\begin{eqnarray}}
\newcommand{\eea}{\end{eqnarray}}
\newcommand{\ba}{\begin{array}}
\newcommand{\ea}{\end{array}}

\begin{document}
\begin{titlepage}
\thispagestyle{empty}

\vspace{2cm}
\begin{center}
\font\titlerm=cmr10 scaled\magstep4 \font\titlei=cmmi10
scaled\magstep4 \font\titleis=cmmi7 scaled\magstep4 {
\Large{\textbf{Holographic energy loss in non-relativistic backgrounds}
\\}}
\vspace{1.5cm}
 \noindent{{Mahdi Atashi $^{}$\footnote{e-mail:ma.atashi@shahroodut.ac.ir}, Kazem Bitaghsir
Fadafan$^{}$\footnote{e-mail:bitaghsir@shahroodut.ac.ir }, Mitra Farahbodnia$^{}$\footnote{e-mail:mitrafarahbod67@yahoo.com}
}}\\
\vspace{0.8cm}
{\em
${}$Physics Department, Shahrood University of Technology, \\ P.O.Box 3619995161 Shahrood, Iran\\}
\vspace*{.25cm}

\vspace*{.4cm}

\end{center}
\vskip 2em

\begin{abstract}
In this paper, we study some aspects of energy loss in non-relativistic theories from holography. We analyze the energy lost by a rotating heavy point particle along a circle of radius $l$ with angular velocity $\omega$ in theories with general dynamical exponent $z$ and hyperscaling violation exponent $\theta$. It is shown that this problem provides a novel perspective on the energy loss in such theories. A general computation at zero and finite temperature is done and it is shown that how the total energy loss rate depends non-trivially on two characteristic exponents $(z,\theta)$. We find that at zero temperature there is a special radius $l_c$ where the energy loss is independent of different values of $(\theta,z)$. Also at zero temperature, there is a crossover between a regime in which the energy loss is dominated by the linear drag force and by the radiation because of the acceleration of the rotating particle. We find that the energy loss of the particle decreases by increasing $\theta$ and $z$. We note that, unlike in the zero temperature, there is no special radius $l_c$ at finite temperature case.

\end{abstract}

\end{titlepage}



\section{Introduction}
From the AdS/CFT correspondence, gravity in the asymptotic AdS geometry is related to the conformal field theory (CFT) on
the boundary. It is well known that the dual quantum field theory would be a strongly coupled theory with a UV fixed point which is invariant under the scaling of space $(\vec{x})$ and time $(t)$. Recently, the generalization to the field theories which are not conformally invariant and the asymptotic geometries are not AdS has been studied. Such field theories are very important in the condensed matter physics. As an example, consider the Lifshitz fixed point theories with the following anisotropic scaling symmetry
\be (t,\vec{x})\rightarrow (w^z t,w\,\vec{x} ),  \ee%
here $w$ is constant and $z$ is called dynamical exponent. The case of $z=1$ is related to the relativistic scale invariance theory. Then one may consider $z\neq1$ theories as non-relativistic theories because of different scaling of time and space.

From the AdS/CFT correspondence, the gravity dual to such theories has been studied in
\cite{Son:2008ye} and \cite{Balasubramanian:2008dm}. From \cite{Kachru:2008yh}, the gravity dual of the Lifshitz fixed point can be found. To derive such non-relativistic geometries from Einstein gravity, one should add other matter fields like massive gauge fields. New backgrounds have been found by including both an abelian gauge field and a scalar field with nontrivial potential as%
\begin{equation}\label{back1}
ds _{d + 2}^2 = {u^{\frac{{2\theta}}{d}}}(\frac{{ - d{t^2}}}{{{u^{2z}}}} + \frac{{d{u^2}}}{{{u^2}}} + \frac{{d{\rho ^2}}}{{{u^2}}} + \frac{{{\rho^2}d{\phi ^2}}}{{{u^2}}} + \frac{{\sum\nolimits_{i = 3}^d {dx_i^2} }}{{{u^2}}}).
\end{equation}
Here, the boundary is located at $u=0$ where $u$ is the bulk or radial direction. Also $\theta$ is hyperscaling violation exponent and the spatial dimension of the boundary field theory is given by $d$. We have written two of the $d$ spatial dimensions using coordinates $(\rho, \phi)$. Physical conditions leads to the following relations between $d$, $\theta$ and $z$ as%
\be (z-1)(d+z-\theta)\geq0,\,\,\,\,\,\,(d-\theta)(d\,z-\theta-d)\geq 0.\ee%

In this paper we probe the non-relativistic theories by studying how the energy loss of a rotating heavy point particle depends on the different values of $\theta$ and $z$. The study of energy loss is an interesting and important problem in studying quark-gluon-plasma (QGP) produced at RHIC and LHC \cite{Matsui:1986dk}. In this case the point particle could be a heavy quark. Study of such problems need non perturbative strongly coupled approaches and time dependent methods, then using the AdS/CFT correspondence is reliable \cite{CasalderreySolana:2011us,DeWolfe:2013cua}. First study of the energy loss of heavy quarks has been done in \cite{Herzog:2006gh,Gubser:2006bz}. They studied a moving heavy quark with a constant velocity through the QGP and calculated the energy required to keep the heavy quark at constant speed. It is found that in this case the energy loss is proportional to the momentum of the moving quark like the energy loss mechanism in the drag force. Then one mechanism for energy loss of the particle comes from the drag force channel. The other possible mechanism for energy loss of the particle could be radiation because of the acceleration of the particle. Study of the energy loss from accelerating objects is basic and interesting problem of quantum field theories. It is very difficult to describe it in the strongly coupled systems. Finding a framework to describe radiation in non-relativistic theories would be very interesting.

Using the AdS/CFT correspondence, we study the energy loss of accelerated heavy objects in the non-relativistic backgrounds. The moving particle is described by a classical string in the bulk. Such string ends on the boundary and the end point corresponds to position of the particle. We study the dynamics of the heavy point particle rotating with a constant angular velocity in the non-relativistic backgrounds with fixed values of $z$ and $\theta$. We assume that the particle rotates along a circle of radius $l$ with a constant angular frequency $\omega$. Then the constant velocity and acceleration are $v=l\omega$ and $a=l^2\omega$, respectively. Here, we discuss how different mechanisms of energy loss, i.e  radiation and drag force could be accessible.

The energy loss of a rotating quark in strongly coupled $N=4$ SYM theory has been first studied in \cite{Fadafan:2008bq}. It was shown that the rotating particle serves as a model system in which two different mechanisms of energy loss is accessible via a classical gravity calculation. This interesting simple model provides a novel perspective on different open questions in studying the energy loss in strongly coupled systems. For example, the radiation pattern of an accelerating quark in a nonabelian gauge theory at strong coupling was studied in \cite{Athanasiou:2010pv}. They find the same angular distribution of radiated power in strongly and weakly coupled regimes which propagates at the speed of light and does not show broadening. Such study at finite temperature plasma has been done in \cite{Chesler:2011nc}. The absence of the broadening is related to the backreaction of the massive particle on the boundary. It means that bulk sources with the speed of light do not generate any energy on the boundary field theory. It was argued in \cite{Hatta:2010dz} that the main reason comes from the supergravity approximation. The expectation value of the energy density sourced by the massive particle with an arbitrary motion was studied in \cite{Agon:2014rda}. Because of the important role of anisotropy during the initial stage of producing QGP, the energy loss of a rotating quark in an anisotropic strongly coupled plasma has also been considered in \cite{Fadafan:2012qu}. For studying of the energy loss of a rotating particle in confining strongly coupled theories see \cite{AliAkbari:2011ue}.

There is an interesting property of energy loss in non-relativistic theories; it was shown that even at zero temperature the drag force in non-relativistic theories such as Schrodinger or Lifshitz theories is not zero \cite{Akhavan:2008ep,Fadafan:2009an,Hartnoll:2009ns}. We have studied before the drag force for asymptotically Lifshitz space times in \cite{Fadafan:2009an}. The drag force on heavy object in an effective theory with hyperscaling violation has been studied in \cite{Alishahiha:2012cm} and \cite{Kiritsis:2012ta}. To explore this effect, one should notice that in the relativistic theories when $z=1$, the energy and momentum are conserved and because of the invariance under boosts the drag force is zero. In $z\neq1$ theories, one finds dissipation and energy and momentum drain into the soft IR modes. It was found that for certain values of $z$ a moving particle only travels a finite distance \cite{Tong:2012nf}. Also this study extended to the case of the response of quantum critical points with the hyperscaling violation to a disturbance caused by a heavy charged particle in \cite{Edalati:2012tc}.

In this paper, we start by studying the energy loss of rotating particle at zero temperature. Surprisingly, we find a critical radius $l_c$ where the total energy loss of the rotating particle does not depend on the non-relativistic parameters of the theory $z$ and $\theta$. At this radius the string end point moves at the speed of light. Also, as we explained unlike the relativistic case, the accelerated heavy  particle looses its energy by drag force. Then one expects that the rate of the energy loss would be the same as the drag force at constant velocity and small acceleration where $a=v\omega\rightarrow 0$, meaning $\omega\rightarrow 0$. One reaches in this limit by increasing the radius $l$ and decreasing the angular velocity $\omega$. We check this behavior numerically by drawing ratio of the total energy loss of the rotating particle to the drag force. Interestingly, plots confirm this statement. Therefore, we study the effect of dynamical $z$ component and hyperscaling violation parameter $\theta$ on this ratio. We summarize the final statements in the discussion section. We also expect the other channel of energy loss,i.e radiation due to acceleration of the particle. In the limit of decreasing $l$ ($l\rightarrow 0$) and increasing $\omega$ ($\omega \rightarrow \infty$), one expects domination of radiation. In the relativistic case, Mikhailov has derived a general result for radiation of an accelerated particle in N=4 SYM vacuum \cite{Mikhailov}. One may describe the radiation by expansion process as the relativistic case and consider a systematic expansion \cite{Tong:2012nf}.\footnote{We would like to thank D. Tong for discussion on radiation in non-relativistic theories.} Although, one can not separate out different effects from radiation by the rotating particle in our problem, it would be useful to understand the physics qualitatively. As it was discussed in \cite{Fadafan:2008bq}, one can not make a sharp distinction between radiation and drag force. We confirm in this study that there are two different regimes at zero temperature as the acceleration-radiation-dominated regime and the drag-dominated
regime, respectively.\\

We extend our study to the case of the energy loss at finite temperature hyperscaling theories. We also note that, unlike in the zero temperature, there is no special radius $l_c$ at finite temperature case.\\

This paper is organized as follows. In section two, we will
present the details of calculation of energy loss at zero temperature and study two different mechanisms of energy loss, i.e drag force and radiation. We study behavior of different mechanisms as a function of parameters of non-relativistic theory $z$ and $\theta$. We consider this study at finite temperature in section three. In the last section we summarize our results. %
\section{Energy loss at zero temperature}
\subsection{Rotating string solution }
In this section, we consider the rotating particle at zero temperature. The background is given by
(\ref{back1}). Based on the AdS/CFT correspondence, the particle is located at the end point of the classical string attached to the boundary of the background geometry. Holographically, the energy loss of rotating quark could be studied by studying the motion of rotating spiral classical string in the geometry (\ref{back1}). The dynamics of the spiraling string governed by the Nambu-Goto action. More details of the calculation can be found in \cite{Fadafan:2012qu}.

The world-sheet ansatz for a rotating quark is paramerized as follows:

\begin{equation}\label{2}
X^{\mu}=\left( t=\tau ,u=\sigma ,\rho=\rho (u),\phi=\omega \tau +\phi (u),x_3=0\right),
\end{equation}
The radial and angular profiles of the rotating string is given by $\rho(u)$ and $\phi(u)$, respectively. They obey the following boundary conditions
\be \label{BC}
\rho(0)=l,\,\,\,\,\phi(0)=0.
\ee

and the Lagrangian density can be obtained as

\begin{equation}\label{Lag}
\mathcal{L} = {u^{\frac{{2\theta }}{d} - 1 - z}}\sqrt{(1 - {\rho^2}{\omega^2}{u^{2(z - 1)}})(1 +
 {{\rho'}^2}) + {\rho^2}{{\phi'}^2}},
\end{equation}
\begin{figure}[ht]
\includegraphics[width=8cm]{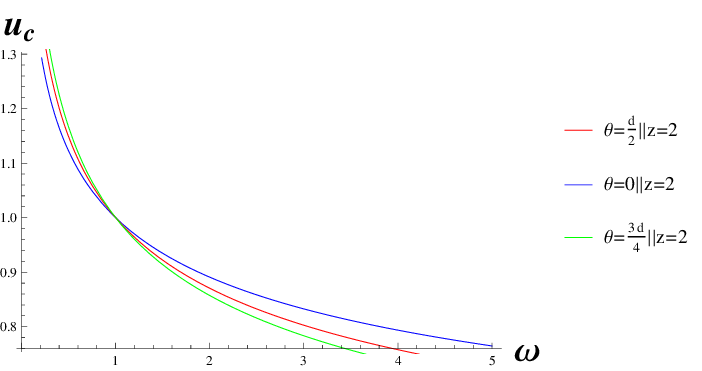} \includegraphics[width=8cm]{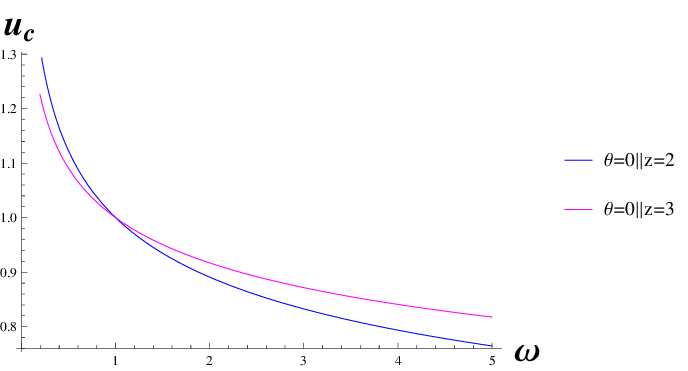} 
\caption{\label{f2-1} Behavior of world sheet horizon $u_c$ vs $\omega$, $\theta$ and $z$  as a function of angular velocity $\omega$. In all plots we fixed $\Pi =1$.}
\end{figure}
where "$^\prime$" shows derivative with respect to "$u$". We then obtain the equations of motion from the Lagrangian. It depends on $\phi'$ but not on $\phi$, so the conjucate momentum, $\Pi_{}$ is a constant of motion and can be written as follows:

\begin{equation}\label{Pi}
{\Pi  } = \frac{{ - \partial L}}{{\partial \phi '}} = \frac{{ - {u^m}{\rho ^2}\phi '}}{{\sqrt {(1 - {\rho ^2}{\omega ^2}{u^n})(1 + {{\rho '}^2}) + {\rho ^2}{{\phi '}^2}} }},
\end{equation}

where $m=\frac{2\theta}{d}-1-z$ and $n=(2z-2)$. Using this constant, the equation of motion for $\rho(u)$ is then given by
\begin{equation}\label{eomT0rho}
\rho'' + (\frac{{\rho (u - m\rho \rho')}}{{u({u^{ - 2m}}{\Pi ^2} - {\rho ^2})}} + \frac{{(2 - n{u^{n - 1}}{\omega ^2}{\rho ^3}\rho')}}{{2\rho (1 - {u^n}{\omega^2}{\rho^2})}})(1 + {\rho'^2}) = 0
\end{equation}

Solving the equation of motion (\ref{eomT0rho}) by considering the boundary conditions in (\ref{BC}), the spiral profile of string will be obtained. By imposing the reality condition on $\phi'$, one finds the special value of the radial coordinate $u_c$. The radius of spiraling string at this value is denoted as $\rho_c$. They can be found as
\be \label{rc}
\rho_c=\frac{\left(\Pi_{}\omega\right)^{\frac{-q}{2m-n}}}{\omega},\,\,\,\,\, u_c=\left(\Pi_{}\omega\right)^{\frac{2}{2m-n}}
\ee

For relativistic case $z=1$, it was shown in \cite{Fadafan:2008bq} that the string in the range of $u<u_c$ which rotates with a speed slower than the local speed of light is casually disconnected from the part of string in $u>u_c$.  While here the local velocity of string is fixed by $v=\rho_c\omega$ and it can change from zero to infinity. This indicates the non-relativistic nature of the dual field theory. In Fig. \ref{f2-1}, we study behavior of $u_c$ in terms of changing angular frequency $\omega$ and dynamical exponents $\theta$ and $z$.

It would be interesting to point out that by rotating the string, a horizon appears on the world sheet and the special point of $u_c$ coincides with a special point on the world-sheet which is denoted as a world sheet horizon. It is very interesting that even at zero temperature it exists while there is no any black hole in the bulk. The physics of the world sheet horizon is related to the Brownian motion of the dual particle in the boundary \cite{Chernicoff:2010yv}. Regarding this study one should consider a gluonic field around the particle in the boundary, also by rotation an "internal degree of freedom is being excited" \cite{Das:2010yw}.

Based on (\ref{rc}), if $\Pi \omega < 1$ or $\Pi \omega > 1$ increasing $z$ and $\theta$ leads to different behavior. As it is clearly seen from Fig. \ref{f2-1}, at fixed $\Pi=1$ the critical value of angular velocity would be $\omega_c=1$. So for $\omega > \omega_c $ effect of increasing $z$ or $\theta$ is not the same as $\omega < \omega_c $. The special point in the bulk direction, $u_c$, is decreasing by increasing $\omega$. Also these two plots show that, there is a critical value for angular velocity, $\omega _c=1$, so that for $\omega <\omega _c$, $u_c$ increases by increasing $\theta$ while decreases by increasing $z$, and vice versa for $\omega >\omega _c$.
\begin{figure}[ht]
\includegraphics[width=8cm]{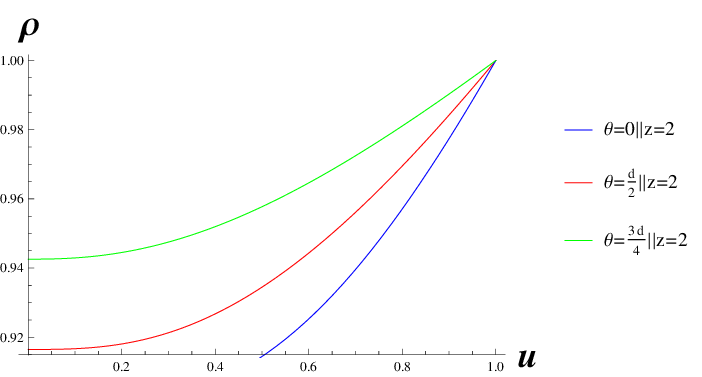} \includegraphics[width=8cm]{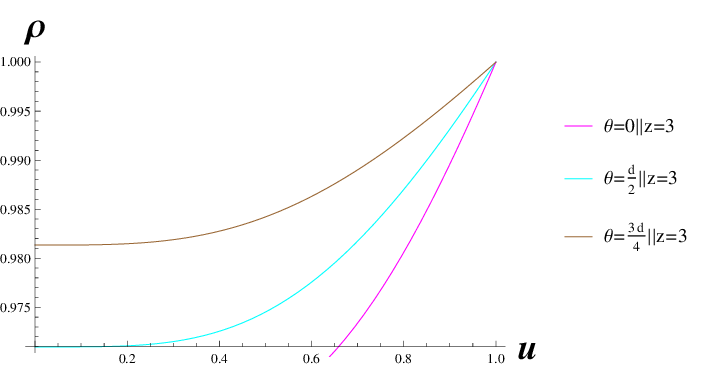} \\ \hfill
\includegraphics[width=8cm]{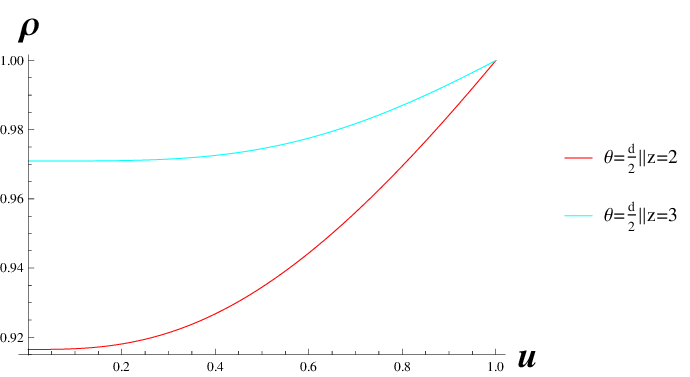} \includegraphics[width=8cm]{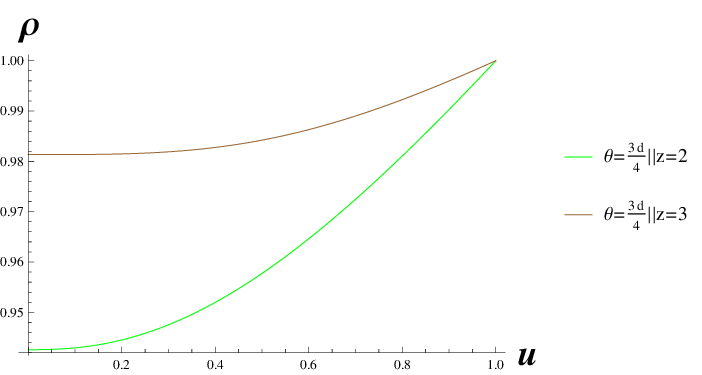} \\ \hfill
\caption{\label{f1} String radius vs radial direction in different $\theta$ and $z$ with $\omega =\Pi =1$. As shown in the top row, by increasing $\theta$, at fixed $u$, the string radius is increasing. As shown in the bottom row, at fixed $\theta$ and $u$, the string radius increases by increasing $z$.}
\end{figure}%

Fig \ref{f1}, shows the behavior of string coiling radius, $\rho (u)$, versus radial direction, $u$, for different values of  $\theta$ and $z$ with $\omega = \Pi = 1$. Notice that we set $u=0$ as boundary, so $\rho (0)$ is the radius of rotation of the particle. The first row of Fig. \ref{f1} shows that at fixed $z$ and $\omega$, the rotation radius for greater value of $\theta$ is bigger. Also the second row shows similar behavior of $\rho$ versus $z$. The generic features of the spiraling string in geometry (\ref{back1}) was discussed in \cite{Alishahiha:2012cm} and the radius of motion $\rho$ as a function of $u$ for the case of $z = 2$ and $\theta=\frac{3d}{4},\frac{d}{2},\,0$ was plotted. 

Briefly, one finds that by increasing $\theta$, at fixed $u$, the string radius is increasing. Also at fixed $\theta$ and $u$, increasing $z$ leads to increasing the string radius.

\subsection{Total Energy loss of rotating particle}

The energy loss of rotating particle is given by

\begin{equation}\label{total}
\frac{dE}{dt}=\frac{1}{2\pi \alpha^{\prime}}\Pi_{} \omega \qquad \qquad
\end{equation}

\begin{figure}[ht]
\includegraphics[width=5cm]{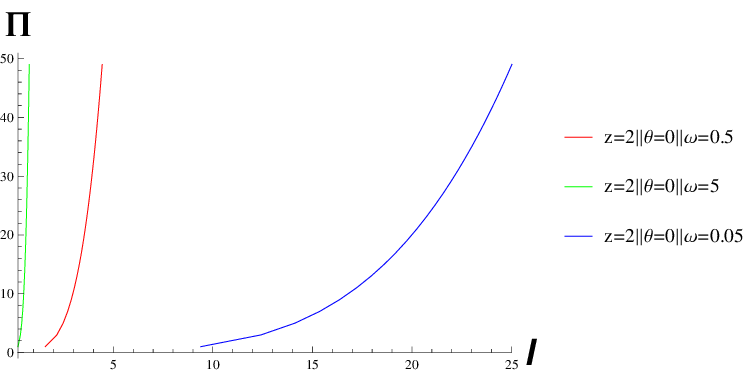} \includegraphics[width=5cm]{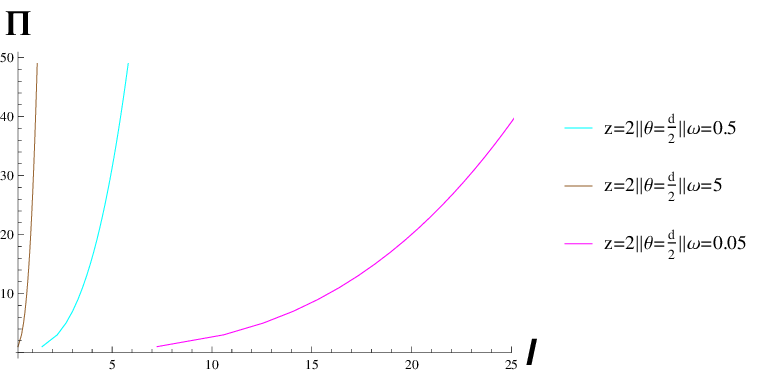}\includegraphics[width=5cm]{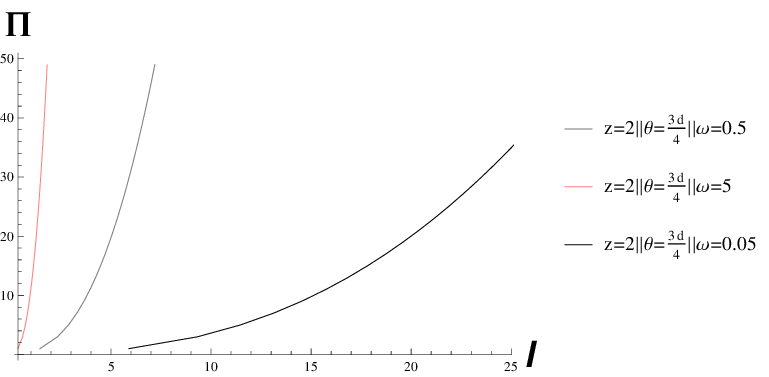} \\ \hfill
\\ \includegraphics[width=5cm]{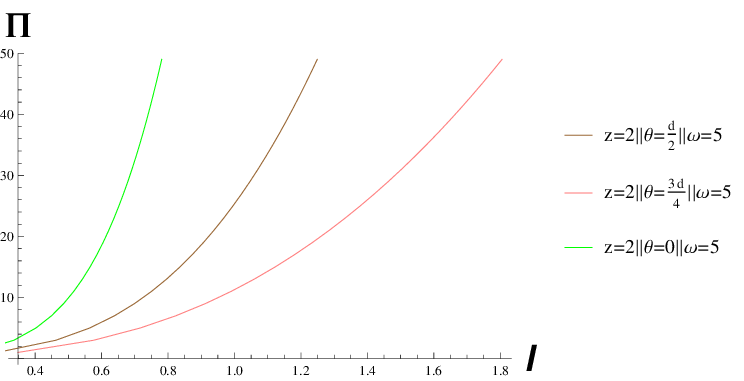} \includegraphics[width=5cm]{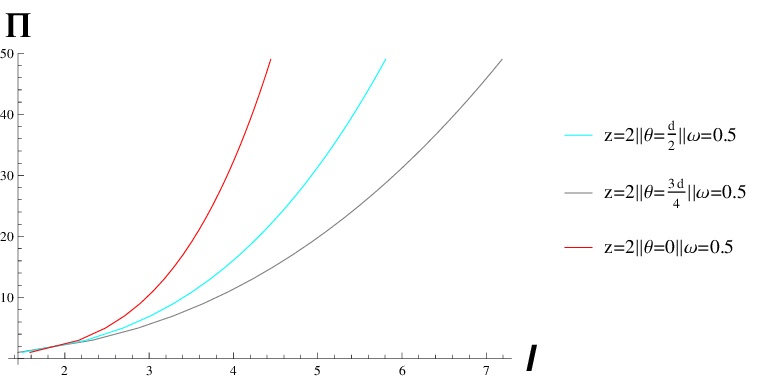}\includegraphics[width=5cm]{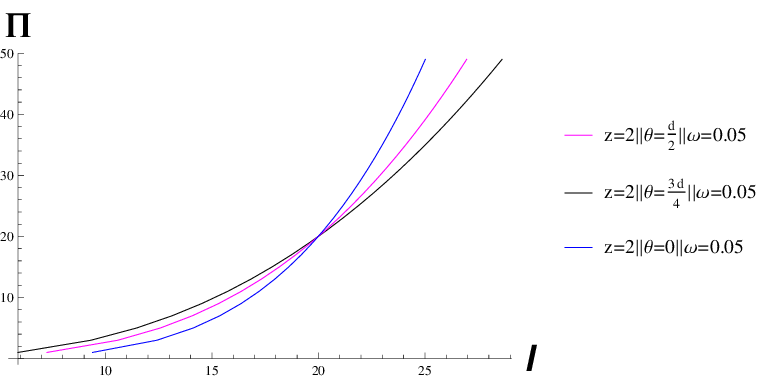} \\ \hfill
\\ \includegraphics[width=5cm]{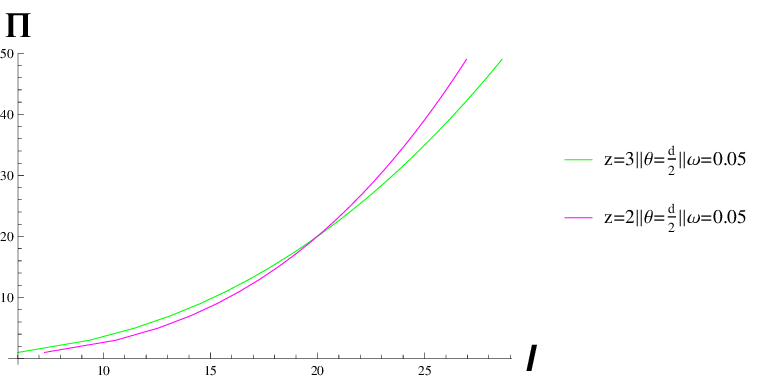} \includegraphics[width=5cm]{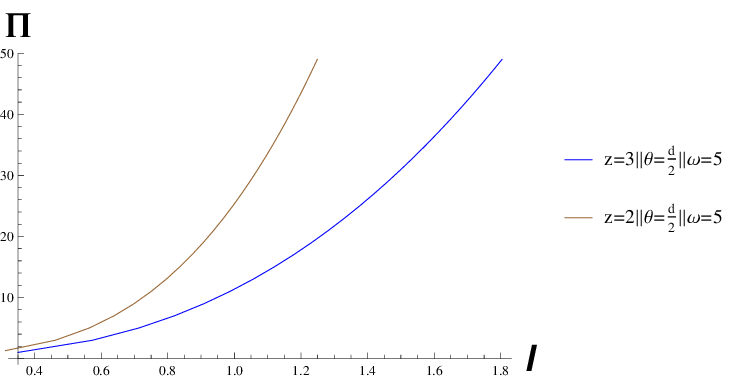}\includegraphics[width=5cm]{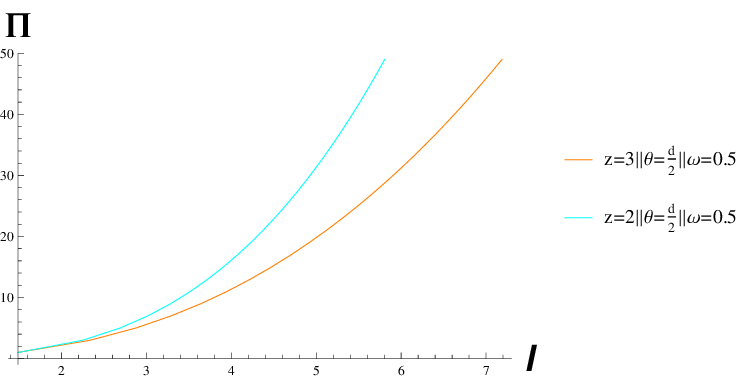} \\ \hfill
\caption{\label{PitoL} Energy loss versus rotation raduis of particle for different values of $\theta$, $\omega$ and $z$. In the first row, we fixed $z=2$ and in each plot from top to down we assumed $\omega=5,\,0.5,\,0.05 $. Also in the first row from left plot to right plot we considered $\theta=0,\,\frac{d}{2},\,\frac{3d}{4}$. In the second row, we fixed $z=2$ and in each plot from top to down we assumed $\theta=0,\,\frac{d}{2},\,\frac{3d}{4}$. Also in the second row from left plot to right plot we considered $\omega=5,\,0.5,\,0.05 $. In each plot of the third row, we fixed $\theta=\frac{d}{2}$ and from left plot to right plot $\omega=0.05,\,5,\,0.5$. }
\end{figure}

We have studied behavior of the energy loss of rotating particle in terms of the radius of the rotation in the boundary in Fig. \ref{PitoL}.  One finds that at fixed $\theta$, by increasing angular velocity $\omega$, the energy loss is increasing for each $z$. Also interestingly, there is a critical value for rotation radius, $l_c$, so that $l_c \omega=1$. For $l>l_c$, at fixed $\omega$ and $l$, the energy loss is increasing by decreasing $\theta$, and vice versa for $l<l_c$. As it is clearly seen from this figure, at $l=l_c$, the energy loss does not depend on $\theta$ and $\omega$. To distinguish between different values of dynamical exponent $z$, we plot the energy loss versus $l$ for $z=2$ and $=3$ in Fig. \ref{PitoL}. One finds that for $l>l_c$, the energy loss corresponding to $z=2$ is greater than energy loss corresponding to $z=3$ in all values for $\theta$ and $\omega$, and vice versa for $l<l_c$. The energy loss is also $z$-independent at $l=l_c$. Briefly, the general features of the energy loss of rotating particle at zero temperature non-relativistic theories can be summarized as

\begin{itemize}
\item As the same as the relativistic case, the energy loss is increasing by increasing of rotation radius $l$.

\item  By increasing rotation velocity $\omega$, the energy loss is increasing for each $z$ and $\theta$.

\item  There is a critical value for rotation radius, $l_c$, so that $l_c \omega=1$ and interestingly the energy loss is $\theta$ and $z$ independent.

\end{itemize}

\begin{figure}[ht]
\includegraphics[width=8cm]{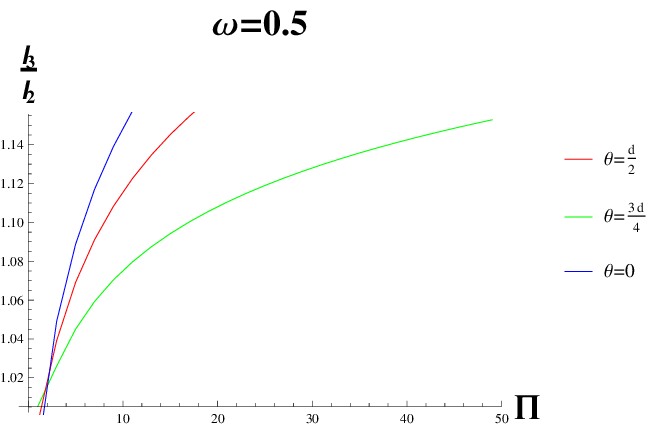} \includegraphics[width=8cm]{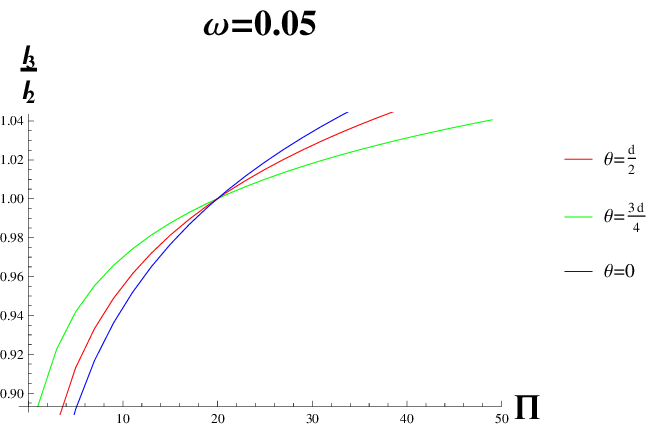} \\ \hfill
\caption{\label{f3} Ratio of $\frac{l_{z=3}}{l_{z=2}}$ vs $\Pi$ for different values of $\theta$ and $\omega$. The point that the ratio is $1$, is the same for similar values of $\omega$ for each value of $\theta$. This point corresponds to $l=l_c$ where the energy loss does not depend on $\theta$, $\omega$ and $z$. }
\end{figure}

To make more clear the role of $z$, we plotted the ratio of $\frac{l_3}{l_2}$ where $l_3$ and $l_2$ are the rotation radius corresponding to $z=3$ and $z=2$, respectively. This ratio is plotted versus energy loss $\Pi$. As shown in Fig. \ref{f3}, where $\Pi \omega =1$ the ratio goes to unity.  This point corresponds to $l=l_c$ where the energy loss does not depend on $\theta$, $\omega$ and $z$.  By increasing of $\theta$ at fixed $\omega$ and $z$, the ratio increases for $\Pi \omega <1$ and decreases for $\Pi \omega >1$.

\subsection{Energy loss of rotating particle from drag force channel}%
\begin{figure}[ht]
\includegraphics[width=8cm]{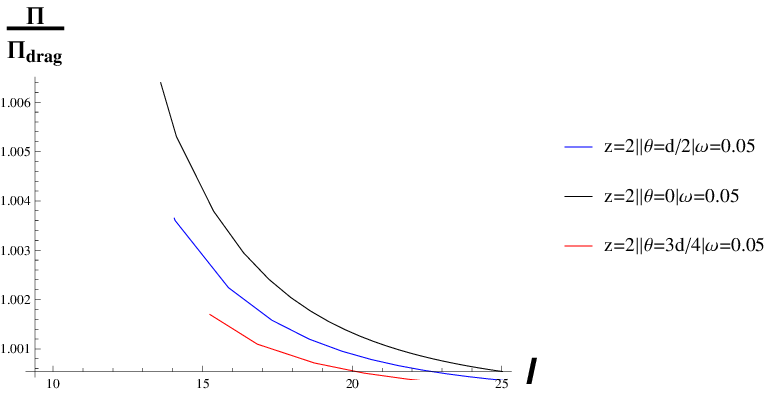} \includegraphics[width=8cm]{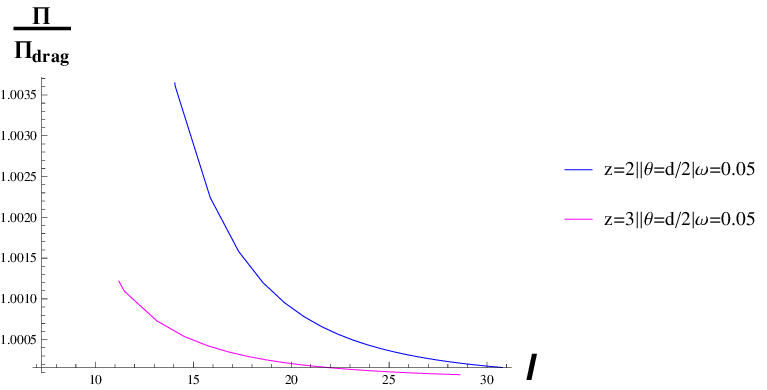}   \\ \hfill
\includegraphics[width=8cm]{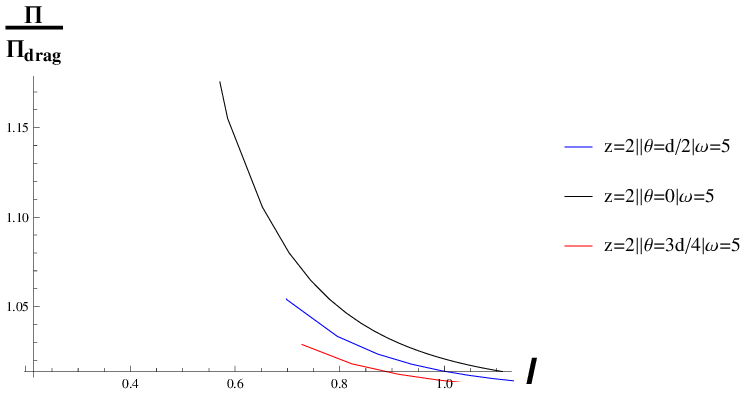} \includegraphics[width=8cm]{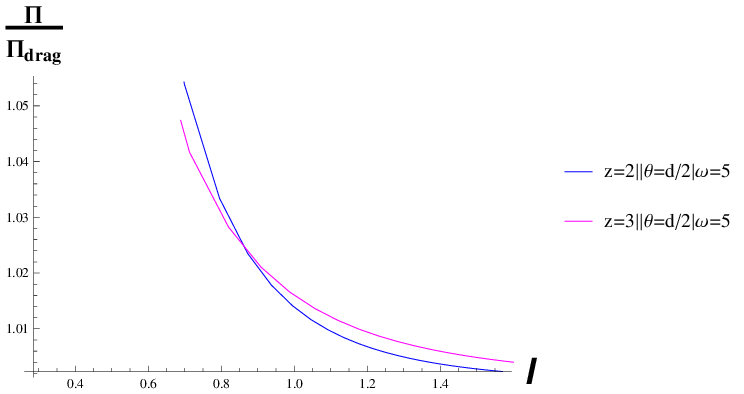}   \\ \hfill
\caption{\label{f4} Ratio of energy loss over drag force vs rotation radius for different values of $\theta$ (left) and $z$ (right) at fixed $\omega=0.05$ and $\omega=5$.}
\end{figure}

To find the drag force, one should consider an open string moving in the non-relativistic background (\ref{back1}). From the correspondence, it represents an external heavy point particle moving with constant velocity $v$ in a non-relativistic field theory. The drag force for
non-relativistic field theories with Lifshitz symmetries or Schrodinger symmetries and theories with
hyperscaling violation have been studied before in \cite{Akhavan:2008ep,Fadafan:2009an,Hartnoll:2009ns,Alishahiha:2012cm,Tong:2012nf,Kiritsis:2012ta}. Interestingly, it was shown that even at zero temperature the energy loss is non-zero which is expressed as
\be\label{dragt0}
2\pi\alpha'\frac{dE}{dt}_{drag}\equiv \Pi_{drag}=v^{\frac{2d-2\theta}{d\,z-d}+2}
\ee
In this subsection, we compare the total energy loss in (\ref{total}) with the energy loss from drag force channel in (\ref{dragt0}). As we discussed in the introduction, one expects that the rate of the energy loss could be the same as (\ref{dragt0}) at constant velocity and small acceleration where $\omega\rightarrow 0$. To reach in this limit, we increase the radius $l$ and decrease the angular velocity $\omega$. We checked this behavior numerically by drawing ratio of the total energy loss of the rotating particle to (\ref{dragt0}) in Fig. \ref{f4}. Interestingly, calculations confirm this statement. For example, one finds from Fig. \ref{f4} that the ratio at fixed $z$ and $\omega$ goes to unity faster for greater value of $\theta$. On the other word at fixed $l$ the ratio corresponding to smaller value of $\theta$ is greater. It means that by increasing $\theta$ the contribution of radiation becomes smaller. Also, the ratio at fixed $\theta$ and $\omega$ goes to unity faster for greater value of $z$. As shown, the ratio corresponding to smaller $z$ is greater at fixed $l$ for $l<l_s$ and vice versa for $l>l_s$, where the $l_s$ is a special rotation radius that the ratio for different value of $z$ are equal there.

\subsection{Energy loss of rotating particle from radiation channel}
One important advantage of studying energy loss of rotating particle is getting information about radiation in strongly coupled non-relativistic theories. As it was discussed before, we expect the other channel of energy loss, i.e radiation due to acceleration of the particle. To dominate this regime, we decrease $l$ and increase $\omega$. For example, see Fig. 5 and consider $\omega=5$ cases. If one fixes the radius of rotating particle, concludes that at fixed dynamical exponent $z$, increasing hyperscaling violation parameter $\theta$ leads to decreasing the energy loss from radiation channel. On the other hand, if one fixes the rate of radiation and increases the parameter $\theta$, finds that $l$ will increase, too. To understand the effect of changing dynamical exponent $z$, we fixed $\theta=\frac{d}{2}$ and increased $z=2$ to $z=3$. It is clearly seen from Fig. \ref{PitoL} that at fixed $l$ increasing $z$ leads to decreasing of radiation of the particle. However, at fixed energy loss by increasing $z$, the radius of rotating particle will increase. It seems that changing $z$ or $\theta$ at fixed $l$ has the same effect on the radiation.

We compare the radiation in non-relativistic theories with relativistic vacuum radiation proposed in \cite{Mikhailov}. In the circular motion, the formula is given by %
\be
\frac{dE}{dt}_{Relativistic}=\frac{\sqrt{\lambda}}{2\pi}\frac{a^2}{(1-v^2)^2}.
\ee
Now we define the following ratio to compare the result with vacuum radiation in $N=4$ SYM theory.
\be \label{radiation-ratio}
\frac{\Pi}{\Pi_{Relativistic}}=\frac{\Pi}{v^2\omega}(1-v^2)^2.
\ee
We plot $\frac{\Pi}{\Pi_{Relativistic}}$ in Fig. 6. This ratio shows that the radiation in strongly coupled non-relativistic theories is smaller than the relativistic case at small velocities. As one expects from (\ref{radiation-ratio}), the ratio vanishes at speed of light, i.e $v=1$. In this figure we fixe $\omega=5$, then at $l=0.2$ the velocity is unity and the ratio goes to zero. We observe that at small velocities of rotating particle the ratio is smaller than unity while by increasing it the ratio increases, monotonically.

\begin{figure}[ht]\label{fradiation}
\centerline{
\includegraphics[width=8cm]{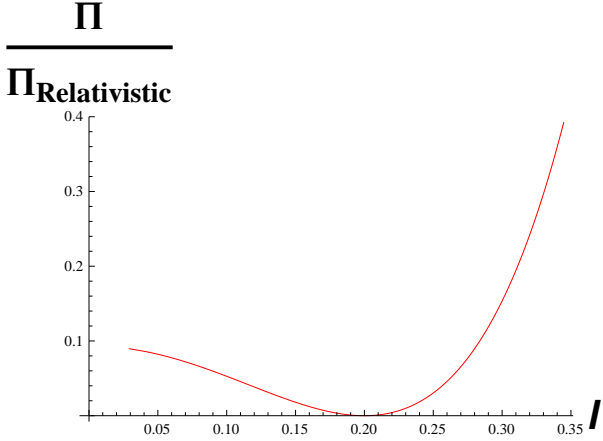}   \\ \hfill
 }
\caption{ Ratio of energy loss over radiation in relativistic vacuum vs rotation radius. }
\end{figure}
\section{Energy loss at finite temperature non-relativistic theory }
So far we have studied the energy loss of particle in quantum field theories with non zero dynamical exponent $z$ and hyperscaling violation $\theta$ at zero temperature. We now generalize the calculations to the case of finite temperature non-relativistic backgrounds. The gravity dual is given by the following metric \cite{Dong:2012se}
\be \label{metric at T}
ds^2=\left(\frac{u}{u_F}\right)^{\frac{2\theta}{d}}\left(-u^{-2z} f(u)dt^2+\frac{du^2}{u^2\,f(u)}+\frac{d\overrightarrow{x}^2}{u^2}\right),\,\,\,\,\,\,
f(u)=1-\left(\frac{u}{u_h}\right)^{z+d-\theta}.
\ee
The boundary is located at $u \rightarrow 0$ and horizon is denoted as $u_h$ which can be  found by solving $f(u_h)=0$. The gravity is valid only for $u>u_F$ which means that $u_F$ is a dynamical scale and we are follwing an effective holographic description. In order to study finite temperature case, one needs to consider $u_F<u_h$. We rescale the metric \eqref{metric at T} by $u_F$ in our analysis. The temperature is given by
\be
T=\frac{d+z-\theta}{4\pi\, u_h^z}
\ee

First, we should use the asatz in \eqref{2} and Nambu-Goto action to study the shape of the spiraling string in the bulk. The lagrangian density is given by
\be
\mathcal{L} = {u^{\frac{{2\theta }}{d} - 1 - z}}\sqrt{\left(f(u) - {\rho^2}{\omega^2}{u^{2(z - 1)}}\right)\left(\frac{1}{f(u)} +
 {{\rho'}^2}\right) + {\rho^2}{{\phi'}^2}f(u)},
\ee
which is exactly matches with \eqref{Lag} when $f(u) \rightarrow 1$.
We assume the constant of motion as $\Pi $, then the equation of motion of $\phi(u)$ is given by
\be \label{phiT}
\phi'^2=\Pi ^2\frac{\left(\frac{-f(u)}{u^{2z}}+\frac{\rho^2\omega^2}{u^2}\right)\left(\frac{1}{u^2f(u)}
+\frac{\rho'^2}{u^2}\right)}{\left(\frac{-\rho(u)^2f(u)}{u^{2+2z}}\right)\left(-\rho(u)^2f(u)u^{\frac{4\theta}{d}-2z-2}
+\Pi^2\right)}
\ee

The world sheet horizon $u_t$ also appears in this case and one finds it by solving the following equations:
\be
\frac{-f(u_t)}{u_t^{2z}}+\frac{\rho_t^2\omega^2}{u_t^2}=0,\,\,\,\,\,\,\,\,\,\,\,\,
-u_t^{\frac{4\theta}{d}-2z-2}\rho(u_t)^2f(u_t)+\Pi^2=0.
\ee

We can use \eqref{phiT} to eliminate $\phi '^2$ from the equation of motion and obtain an equation of motion for $\rho(u)$ in terms of the constant $\Pi $. The differential equation after the partial decomposition can be written as the following form
\begin{align}\label{ODET}
&\rho''(u)+\frac{u^{\frac{4\theta}{d}-1}\rho(u)\left( -2d\,u\,+\rho(u)\left(-2f(u)(d+dz-2\theta)+duf'(u)\right)\rho'(u) \right)\left(1+f(u)\rho'(u)^2\right) }{2\,d\,\rho(u)\left(-\Pi^2\,u^{2+2z}+u^{\frac{4\theta}{d}}f(u)\rho(u)^2 \right)}+\nonumber\\
&\frac{2u^3+\rho'(u)\left(u^3\rho(u)h'(u)+2 u^3 f(u)\rho'(u)+u^{2\,z}\omega^2\rho(u)^3\left(2-2z+2f(u)-2zf(u)+uf'(u)\right)\rho'(u)^2\right)}{2u\,\rho(u)\left(u^2\,f(u)-u^{2\,z}\,\omega^2\,
\rho(u)^2\right)}=0.
\end{align}
This is an important result as general equation of motion for the rotating particle in the finite temperature non-relativistic theories whith different values of $\left(T,d,z,\theta\right)$. The same general equation at zero temperature was given in \cite{Alishahiha:2012cm}.

One may check this result by assuming $\left(d=3,\theta=0,z=1,f(u)\rightarrow 1\right)$ and finding the differential equation for the rotating particle at $N=4$ SYM theory as \cite{Fadafan:2008bq}
\be
\rho''(u)+\frac{\rho(u)(1+\rho'(u)^2)\left(u+2\rho(u)\rho'(u)\right)}{u\left(\Pi^2 u^4-\rho(u)^2\right)}
+\frac{1+\rho'(u)^2}{\rho(u)\left(1-\omega^2\rho(u)^2\right)}=0.
\ee
Also at finite temparure $N=4$ SYM theory with $\left(d=3,\theta=0,z=1\right)$, one finds the same result as \cite{Athanasiou:2010pv,Fadafan:2008bq}
\begin{align}
&\rho''(u)+\frac{\rho(u)\left(1+f(u)\rho'(u)^2\right)\left(4f(u)\rho(u)\rho'(u)+u\left(
2-\rho(u)f(u)\rho'(u)\right)\right)}{2u\left(-\Pi^2u^4+f(u)\rho(u)^2\right)}+\nonumber\\
&\frac{2+\rho(u)f'(u)\rho'(u)+2f(u)\rho'(u)^2+\omega^2\rho(u)^3f'(u)\rho'(u)^3}{2\rho(u)\left(f(u)-\rho(u)^2\omega^2\right)}=0.
\end{align}
One can use the special point $(\rho_t,u_t)$ as initial value for the differential equation of motion and solve it. The second initial condition $\rho'_t$ can be found from the prescription of \cite{Fadafan:2008bq} by using an expansion of $\rho(u)$ around $u=u_t$. One finds that the differential equation itself determines $\rho'_t$, one  obtains $\rho'_t$ by solving the following equation:
\begin{align}
&-2\sqrt{f(u_t)}u_t^z(d-\theta)\omega+2 f(u_t)^{\frac{3}{2}}u_t^z\,(d-\theta)\omega \,\,\rho'(u_t)^2+\nonumber\\
&\rho'(u_t)\left(\theta\,f(u_t)\left( f'(u_t)u_t-4f(u_t)(z-1) \right)
+d\left(-zu_tf(u_t)f'(u_t)+2f(u_t)^2(z^2-1)+2u_t^{2z}\omega^2\right)\right)=0.
\end{align}
\begin{figure}[ht]
\includegraphics[width=8cm]{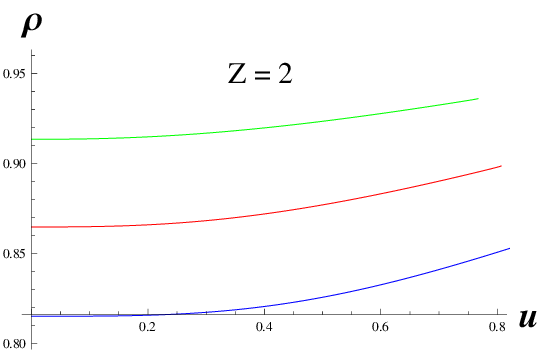} \includegraphics[width=8cm]{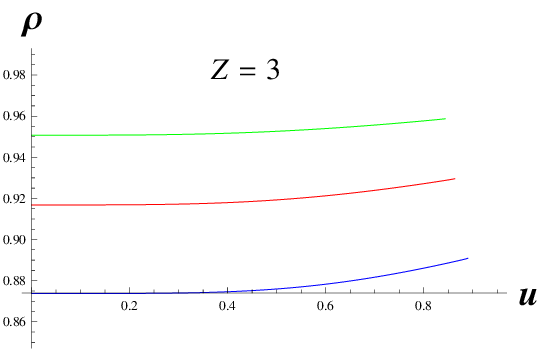} \\ \hfill
\caption{\label{Ltz} The finite temperature analyziz of the string radius vs radial direction in different $\theta$ and $z$ with $\omega =\Pi =1$. In each plot from top to down $\theta=\frac{3d}{4},\,\frac{d}{2},\,0.$}
\end{figure}%

As it is clear this is a general equation which depends on the parameters of the non-relativistic theory.
In the case of $N=4$ SYM theory, it reduces to
\be
-1+u_t\,\omega\,\rho'(u_t)+\rho'(u_t)^2=0.
\ee
For finite temperature $N=4$ SYM theory, one finds
\be
-2\sqrt{f(u_t)}\omega+\left(-f(u_t)f'(u_t)+2u_t \,\omega^2\right)\rho'(u_t)+2 f(u_t)^{3/2}\,\omega\,\rho(u_t)^2=0.
\ee
\\\\
Having initial values of $\rho(u_t)$ and $\rho'(u_t)$, one can solve \eqref{ODET} for different values of $\left(T,d,z,\theta\right)$. We fixe the temperature in our analyziz and consider $u_h=1.$ Also assume $d=3$. Therefore, study how changing of $\theta$ and $z$ affect the energy loss of the rotating particle.\\

First like Fig \ref{f1} we study radial string configuration for the rotating particle at finite temperature in Fig \ref{Ltz}. This figure shows the behavior of string coiling radius, $\rho (u)$, versus radial direction, $u$, for different values of  $\theta$ and $z$ with $\omega = \Pi = 1$ at finite temperature. Again we set $u=0$ as boundary, so $\rho (0)$ is the radius of rotation of the particle. In each plot from top to down $\theta=\frac{3d}{4},\,\frac{d}{2},\,0.$ One finds that at fixed $z$ and $\omega$, the radius of the particle increases by turning on $\theta$, i.e increasing $\theta$ leads to increasing $l$, significantly. At fixed $\theta$ and $\omega$, by increasing $z$ the radius also increases. Also the string does not bend outward by increasing $z$. Then it is clear that the shape of spiraling string depends on the values of non-relativist theory, significantly.

\begin{figure}[ht]\label{PiLT}
\includegraphics[width=5cm]{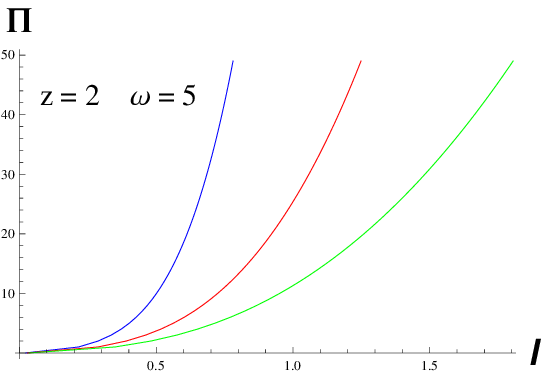} \includegraphics[width=5cm]{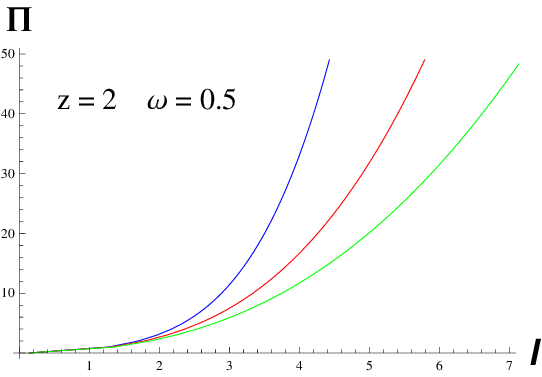}\includegraphics[width=5cm]{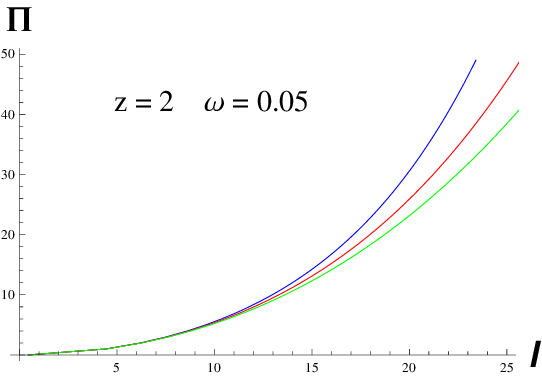} \\ \hfill
\\ \includegraphics[width=5cm]{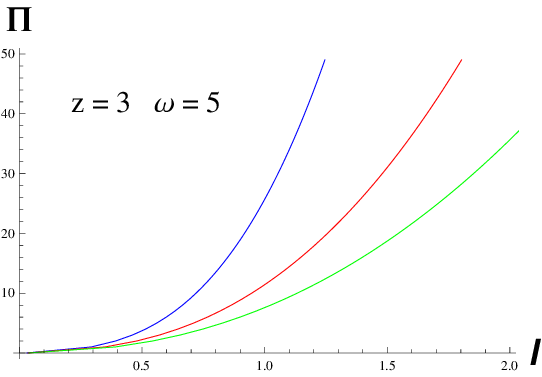} \includegraphics[width=5cm]{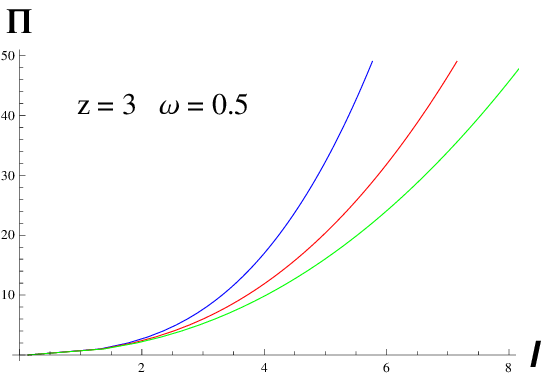}\includegraphics[width=5cm]{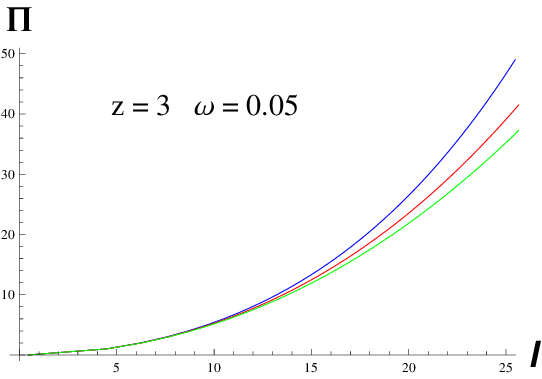} \\ \hfill
\caption{\label{f2} Energy loss versus rotation raduis of the rottaing particle for different values of $\theta$, $\omega$ and $z$ at finite temperature. In the first row we consider $z=2$ and change the angular velocity from left plot to right plot $\omega=5,\,0.5,\,0.05$. In the second row we increased dynamical exponent $z$ to $z=3$. In each plot from top to down $\theta=0,\,\frac{d}{2},\,\frac{3d}{4}$.}
\end{figure}

Next we study the energy loss of the test particle at finite temperaure. The energy loss is given in terms of the the constant $\Pi$ as
\be
\frac{dE}{dt}=\frac{1}{2\pi \alpha'}\Pi\,\, \omega.
\ee
Then we choose different values for $(z,\theta)$ and find the energy loss of particle at finite temperature.  In this way, we should study behavior of $u_t$ from (18) and find shape of the spiraling string in the bulk by solving \eqref{ODET}. Here, we consider $\theta=0,\,\frac{d}{2},\,\frac{3d}{4}$ and change values of dynamical exponent $z$. Fig. 8 presents the energy loss at different values of the angular frequency $\omega$. As it is clear from this figure the energy loss monotically increases by increasing the radius of the circular rotation. Also by increasing $\omega$, radius $l$ decreses significantly at fixed $\Pi$. In particular, for large values of $\omega$ only small values of $l$ is allowded. We also studied the effect of increasing $z$ on the enrgy loss in the second row of this figure. An important result is that at at fixed value of $l$, increasing $z$ leads to decreasing of the enrgy loss of rotating particle. Increasing hyperscaling violation parameter $\theta$ also leads to decreasing of the energy loss of the particle. Briefly, the general features of the energy loss of rotating particle at finite temperature non-relativistic theories can be summarized as
\begin{itemize}
\item As the same as the zero temparature non-relativistic theory, the energy loss is increasing by increasing of rotation radius $l$, monotonically.

\item  By increasing $z$ and $\theta$, the energy loss of the rotating particle decreses.

\item  Unlike the zero temparature non-relativistic theory, there is a critical value for rotation radius, $l_c$.
\end{itemize}
We checked the last staement by considering different values of $z$ and $\theta$, however, it seems that there is no such critical radius at finite temperature non-relativistic theory.
\textbf{Linear drag force limit,} Because of moving the particle at finite temperaure strongly coupled field theory, one expects that the particle experiences the linear drag force. Such calculation has been done in \cite{Akhavan:2008ep,Fadafan:2009an}. Now we want to  study if the same mechanism exists for the total energy loss of the particle, i.e it looses the energy due to the linear drag force.

One finds the detail of drag force calculations in \cite{Fadafan:2009an}. The rate of the energy loss of heavy probe is given by
\be
P \equiv 2\pi\alpha' \frac{dE}{dt}=v^2\,u_c^{-2+\frac{2\theta}{d}},
\ee
where $u_c$ should be found by solving $f(u_c)u_c^{-2z+2}-v^2=0.$

\begin{figure}[ht]
\centerline{
\includegraphics[width=5cm]{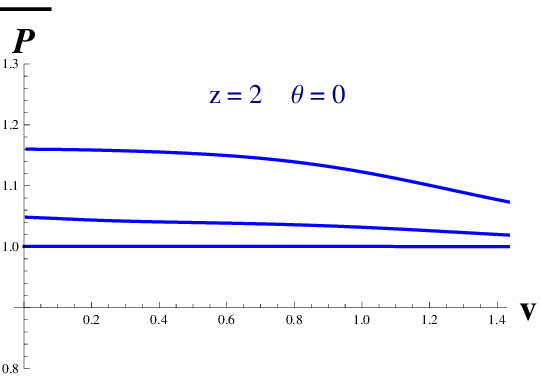} \includegraphics[width=5cm]{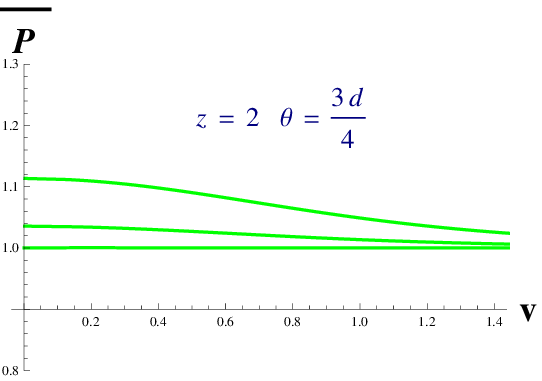}\includegraphics[width=5cm]{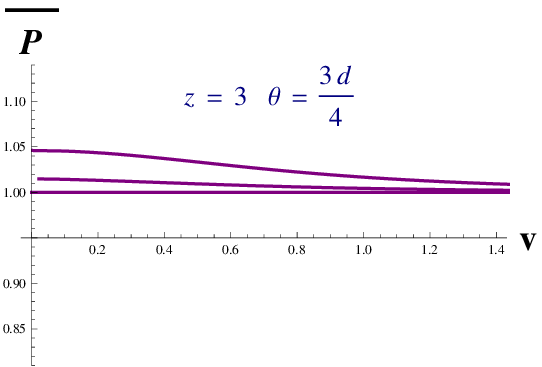} \\ \hfill
 }%
\caption{\label{PitoP} The ratio of the total energy loss of rotating particle to the linear drag force at finite temperature as a function of velocity. In each plot from top to down $\omega=1,\,0.5,\,0.05$.}
\end{figure}

Comparing with $z=1$ case in \cite{Fadafan:2008bq}, N=4 SYM at finite temperature, at small angular velocity $\omega$, where $a\rightarrow 0$, the rotating particle looses its energy from linear drag force on a particle moving in a straight line with a constant velocity $v$ at finite temperature.
Here, we show the ratio of the total energy loss of rotating particle to the linear drag force at finite temperature as a function of velocity in the Fig. \ref{PitoP}. Because of the non-relativistic theory the velocity can change from zero to infinity. In each plot of this figure, from top to down the angular velocity $\omega$ decreases as $\omega=1,\,0.5,\,0.05$. We have changed $z$ and $\theta$ to study the ratio. It is clearly seen that the enrgy loss from the linear drag force mechanism is dominant when $\omega$ goes to zero. Increasing the parameters of $z$ and $\theta$ also leads to this phenomena. However, the total energy loss exceeds the linear drag by increasing the angular velocity $\omega$ or decreasing the parameters of $z$ and $\theta$.

\section{Discussion}
In this paper, we have studied some aspects of energy loss in non-relativistic theories from holography. It has been done by analyzing the energy lost by a rotating heavy point particle along a circle of radius $l$ with angular velocity $\omega$ in holographic non-relativistic field theories with general dynamical exponent $z$ and hyperscaling violation exponent $\theta$. It was shown that this problem provides a novel prespective on the energy loss in non-relativistic strongly coupled field theories. A general computation at zero and finite temperature is done and it is shown that how the total energy loss rate depends non-trivially on two characteristic exponents, $(z,\theta)$. One should notice that there is a null singularity in such theories which could be resolved by adding some stringy corrections. We have ignored such issues and considered only some special range of the energies.

First, by studying the shape of the spiraling string at zero and finite temperature we showed that a world-sheet horizon appears at $u_c$. From the boundary theory point of view, one can distinguish between the far field region or the near field region, too. Also, as the particle radiates it should experience small kicks leading to Brownian motion of the particle in no-relativistic theories. We have shown how the world sheet horizon $u_c$ depends on the $(\theta,z)$ in Fig. 1.

Next, we have studied $\frac{dE}{dt}$, the energy lost by the particle, which is the energy expended by the external force moving the particle. The calculation has been done at zero and finite temperature. This problem has been studied in N=4 SYM and although it is not a physical situation, however novel preprctives were found by studying it. It was shown that at zero temperature there is a special radius $l_c$ where the energy loss is independent of different values of $(\theta,z)$. It was shwon in \cite{Tong:2012nf,Edalati:2012tc} that there is a qualitative difference between non-relativistic theories with different values of $z$. For example, if one throws a massive particle in one of the boundary theory directions it will travel only a finite distance in the case of $z=2$ theory \cite{Tong:2012nf}. Following these studies, we expect to find different behaviors for rotating particle by changing $(\theta,z)$.  We summarize our results at as

\begin{itemize}
\item We found that like the relativistic case, the energy loss is increasing by increasing of rotation radius $l$.

\item It was shown that by increasing angular velocity $\omega$, the energy loss is increasing for different values of $z$ and $\theta$.

\item We found that at zero temperature there is a critical value for rotation radius, $l_c$, so that $l_c \omega=1$ and interestingly the energy loss is independent of parameters of $\theta$ and $z$.
\item  By increasing $z$ and $\theta$, the energy loss of the rotating particle decreses.

\item  At finite temparature non-relativistic theory, there is no a critical value for rotation radius, $l_c$.
\item  At zero and finite temeparur, it was shown that, like N=4 SYM case at finite temperature, there is a crossover between a regime in which the energy loss is dominated by the linear drag force and by the radiation because of the acceleration $a$. However, the radiation process in such theories with non zero $(z,\theta)$ is not the same as the Mikhailov result proposed in \cite{Mikhailov}.
\end{itemize}

It is worth to remark again that the linear drag force even at zero temperature is not zero. Therefor, two different mechanisms of linear drag force would be linear drag force at zero and finite temperature. We analysed the total energy loss at finite temperature by studing its ratio to the drag force at zero and finite temperature and we could observe the limit where the drag force channel dominates. Then we saw the significant advantage offered by the analysis of the rotating object in the non-relativistic theories.

Another feature of the energy loss of the particle at finite temperature that is worth emphasizing is to study the energy loss due to linear drag at zero temperature. It means that if particle looses its energy from drag force channel at zero temperature in \eqref{dragt0}. In this case, we considered different values of $z$ and $\theta$ and studied ratio of the total energy loss, i.e $\Pi\omega$ to the drag force at zero temperature. To better underestand the physics, we studied also the ratio of the total energy loss to the linear drag force at zero plus the linear drag force at finite temperature. By changing $z$ and $\theta$, we could find that when the partcle rotates very fast this ratio goes to one. This means that the particle looses its energy from the linear drag force mechanism at zero and finite temperature, i.e theay have almost same contribution in the energy loss of the particle.

Based on holography, the rotating particle has a gravitational dual; a semiclassical string. The end point of this string corresponds to the particle in the non-relativistic quantum field theory. So the rotating motion of particle corresponds to coiling of the classical string about the same axis of the rotation. From the correspondence, the string induces a $4d$ stress tensor on the boundary that is expectation value of stress tensor of the boundary field theory. As a result, we can study quantum effects in the boundary field theory by classical calculation in the bulk. Thus, one can use the qualitative idea behind the AdS/CFT correspondence and study how the classical depths of the spiraling strings in the bulk is related to the length-scale in the non-relativistic theory. Because that depth in the radial direction corresponds to length-scale in the field theory. Such study has been done in the case of N=4 SYM theory in \cite{Chesler:2011nc}. Surprisingly, they found that this intuitive result from the AdS/CFT correspondenceway does not work and the rotating string falls deeper and deeper into the radial direction while the thickness of the flux tube of energy density in the boundar field theory does not change. It would be very interesting to check if such behavior also exist in our case, we leave this study to further work.


\end{document}